\documentstyle[prl,aps,epsfig]{revtex}

\newcommand{\KET}[1]{{| #1 \rangle}}

\newcommand{\cuoq}[0]{\mathrm{CuO}_{4} }

\newcommand{\commento}[1]{}

\newcommand{\bgst}[2]{{\langle \Psi_{0}^{(#1)}(#2) |}}
\newcommand{\kst}[3]{{| \Psi_{#3}^{(#1)}(#2) \rangle}}

\newcommand{\be}{\begin{equation}}
\newcommand{\ee}{\end{equation}}
\newcommand{\beq}{\begin{eqnarray}}
\newcommand{\eeq}{\end{eqnarray}}

\newcommand{\cu}[0]{\mathrm{CuO}_{4} }

\begin{document}
   
\def\gC{\mbox{\boldmath $C$}}
\def\gZ{\mbox{\boldmath $Z$}}
\def\gR{\mbox{\boldmath $R$}}
\def\gN{\mbox{\boldmath $N$}}
\def\ua{\uparrow}
\def\da{\downarrow}
\def\a{\alpha}
\def\b{\beta}
\def\g{\gamma}
\def\G{\Gamma}
\def\d{\delta}
\def\D{\Delta}
\def\e{\epsilon}
\def\ve{\varepsilon}
\def\z{\zeta}
\def\h{\eta}
\def\th{\theta}
\def\k{\kappa}
\def\l{\lambda}
\def\L{\Lambda}
\def\m{\mu}
\def\n{\nu}
\def\x{\xi}
\def\X{\Xi}
\def\p{\pi}
\def\P{\Pi}
\def\r{\rho}
\def\s{\sigma}
\def\S{\Sigma}
\def\t{\tau}
\def\f{\phi}
\def\vf{\varphi}
\def\F{\Phi}
\def\c{\chi}
\def\w{\omega}
\def\W{\Omega}
\def\Q{\Psi}
\def\q{\psi}
\def\de{\partial}
\def\inf{\infty}
\def\ra{\rightarrow}
\def\bra{\langle}
\def\ket{\rangle}

\draft

\twocolumn[\hsize\textwidth\columnwidth\hsize\csname
@twocolumnfalse\endcsname

\widetext

\title{Symmetric Hubbard Systems with Superconducting Magnetic
Response}

\author{Agnese Callegari, Michele Cini,  Enrico Perfetto and Gianluca
Stefanucci}
\address{Istituto Nazionale per la Fisica della Materia, Dipartimento di
Fisica,\\
Universita' di Roma Tor Vergata, Via della Ricerca Scientifica, 1-00133\\
Roma, Italy}

\maketitle

\begin{abstract}

In purely repulsive, $C_{4v}$-symmetric Hubbard clusters a
correlation effect produces an effective two-body attraction and pairing;
the key ingredient is the availability of  $W=0$ pairs, that is,
two-body solutions 
of appropriate symmetry. We study the tunneling of bound pairs in rings of
5-site units 
connected by weak intercell links; each unit has the topology of a
CuO$_{4}$ cluster and a repulsive interaction is included on every site.
Further,  we test the superconducting
nature of the response of this model to a threading magnetic field.
We present a  detailed numerical study of the two-unit ring
filled with 6 particles and the three-unit ring with 8 particles;
in both cases a lower filling yields normal behavior.
In previous studies on $1d$  Hubbard chains,
level crossings were reported (half-integer or fractional
Aharonov-Bohm effect) which however cannot be due to superconducting
pairs. In contrast, the nontrivial basis of clusters carrying $W=0$ pairs
leads to genuine Superconducting Flux Quantization (SFQ).
The data are understood in terms of  a
cell-perturbation theory  scheme which is  very accurate for weak
links. This low-energy approach leads to an effective
hard core boson Hamiltonian which naturally  describes itinerant
pairs and SFQ in mesoscopic rings.
For the numerical calculations, we take advantage of  a recently
proposed exact diagonalization technique which can be generally applied
to many-fermion problems and
drastically reduces the size of the matrices to be handled.

\end{abstract}

\bigskip\bigskip\bigskip
]

\narrowtext
\section{Introduction}

{\small 

Low-dimensional systems with strong electron-electron correlations may
lead to an anomalous Aharonov-Bohm (AB) effect\cite{ba} with ground-state
energy 
oscillations versus flux $\phi$ having a period shorter than the
foundamental one, 
given by $\f_{0}=hc/e$. However, as we shall see below, a fractional
period $\f_{0}/N$ does not generally mean that the current is carried by
particles with an effective charge $e^{\ast}=Ne$. In particular, if
$N=2$ the half-integer AB effect is not equivalent to the
superconducting flux quantization (SFQ) and pairing is not
necessarily implied.

The repulsive Hubbard ring in the presence of a magnetic flux has
been studied by several authors. At half filling it was
inferred by numerical evidence\cite{scala2} that the period
of the ground state energy $E^{(0)}(\f)$ as a function of the
trapped flux $\f$ is a whole fluxon if $U/t$ is in a physical parameter
range. Away from half filling Kusmartsev\cite{kusmartsev1} pointed out that
for microscopic Hubbard rings with $N$ particles the ground state
energy has $N$ minima in the range $[0,\f_{0})$ for $U=\inf$ and at low
density. [This result has been confirmed by Schofield {\em et al.} in
Ref.\cite{schofield}.] These oscillations were originally explained
in terms of spin-flip processes by the same author, so that the system
varies its total spin  as the flux changes by
one flux quantum. However, when $U=\inf$ many degenerate ground states  with
different total spin $S$ exist, and a more accurate explanation of the
fractional AB effect has been provided by Yu and
Fowler\cite{yufowler}\cite{kusmartsev2}.
They studied the Lieb-Wu\cite{liebwu} equations for a chain with twisted
boundary 
conditions [a flux $\f$ corresponds to a twist of $2\p\f/\f_{0}$] with
leading $t/U$ corrections. Increasing the flux the holon momenta get
shifted and the energy of the holon sea grows; to
counterbalance this effect the system generates a compensating
momentum by creating a hole in the distribution of the spinon quantum
numbers. This excitation in the spinon sea is not energetically
suppressed as far as $Nt/LU\ll 1$ where $L$ is the number of sites.
The quantization of the spinon
momenta does not allow a full compensation for the effect of a
continuosly varying flux $\f$ and leads to the energy oscillations
discussed above. As $Nt/LU$ increases the spinon excitation
energies are raised and beyond a critical value $E_{0}(\f)$ has only
two minima at $\f=0$ and $\f=\f_{0}/2$ corresponding to the spinon
excitations with momentum 0 and $\p$, as numerically observed
in Refs.\cite{scala2}\cite{yufowler}\cite{ferretti1}. Hence, the
half-integer AB effect is driven by level-crossing due to the spinon
degrees of freedom and it is not SFQ.

Recently, an exact result on the infinite $U$ Hubbard ring with
twisted boundary conditions has been obtained by Nakano\cite{nakano}.
He proved that the ground state energy $E^{(0)}(\phi)$ with an even
number of particles $N$ is periodic with period $\phi_{0}/2$
if the $z$-component of the total spin $S_{z}=0$ and with period
$\phi_{0}/N$ if $M_{\ua}/M_{\da}$ (with
$M_{\ua}\geq M_{\da}$) is not an integer; here $M_{\s}$ is the number
of particles with spin $\s=\ua,\da$. 
We emphasize that the Nakano theorem is not in contrast 
with the results by Kusmartsev. Indeed, even if 
$E^{(0)}(\phi)$ has period $\phi_{0}/2$ it may exhibit 
other minima in the range $[0,\phi_{0})$.

The half-integer AB effect caused by the existence of
confined pairs has been observed in the framework of extended 1$d$ Hubbard
models. Sudb$\o$ {\em et al.}\cite{sudbo} studied an alternating Cu-O
ring with a charge-gap between Cu and O (which
favours the particles to occupy O sites) and an on-site repulsion is allowed
on Cu's but not on O's. In this way a half-integer flux quantization
needs  a strong  off-site repulsion; pair confinement is achieved by
allowing for repulsion-free sites, and not by producing  an effective
attraction. 
Other authors considered modified 1$d$ models with bond-charge
interactions so that the number of doubly occupied sites is
an extra conserved quantity\cite{arrachea1}. Turning on a term
breaking the extra symmetry gives rise to an effective attractive
interaction and to the SFQ for moderate
on-site repulsion. However, the pairing mechanism is not driven by
the Hubbard interaction as witnessed by the presence of SFQ
even at $U=0$\cite{arrachea2}\cite{ferretti2}.

Few data are avaliable in $2d$ Hubbard systems, since no exact solutions
exist in this case and numerical explorations are possible only for small
clusters. Long ago Canright and Girvin\cite{canright} discussed the 
magnetic response of the  Hubbard model showing SFQ
in the {\em attractive} case by threading a magnetic field into a
cylindrical probe.
Assaad {\em et al.}\cite{ahs} computed numerically the magnetic response of
the repulsive Hubbard model in the same geometry at quarter
filling, but the flux quantization was found to be normal. 
Signatures of anomalous flux quantization in a $4\times 4$ geometry 
have been provided by Arrachea {\em et al.}\cite{arrachea3}; however 
they studied an {\em extended} Hubbard model with nearest-neighbour 
correlated, e.g. occupation-dependent, hopping.

All these results would suggest that the half-integer AB effect
cannot be interpreted as SFQ in the context of repulsive Hubbard systems.
However, Anderson\cite{anderson} first
advocated the non-conventional superconductivity arising from
repulsive interactions, proposing the one-band Hubbard Hamiltonian as a
prototype model.
Evidence for pairing  has then been
obtained by several authors by a variety of methods.
Analytic approaches based on a renormalization
method\cite{EPJB1999}\cite{SSC1999} and
on various implementations of the renormalization
group technique\cite{zanchi}\cite{halboth}\cite{honerkamp},
generalized conserving approximation theories
like FLEX\cite{flex}, as well as  Quantum Monte  Carlo studies on
supercells\cite{fettes} lead to this conclusion. In
Refs.\cite{cibal1}\cite{cibal2}\cite{cibal3}\cite{cibal4}
we show that pairing occurs also in purely repulsive, $C_{4v}$-symmetric
Hubbard clusters.

In this paper we wish to study the tunneling
of bound pairs in rings of 5-site units with a CuO$_{4}$ topology, see 
Fig.(\ref{topo}). 
\begin{figure}[H]
\begin{center}
   \epsfig{figure=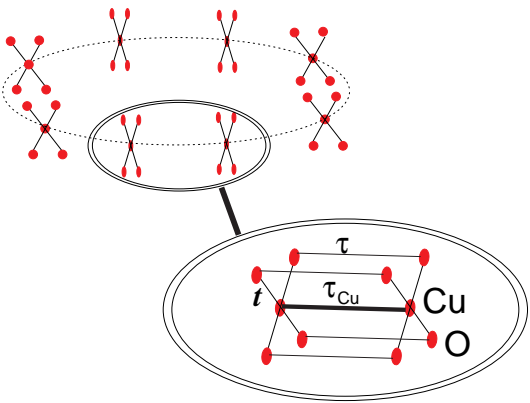,width=7cm}\caption{\footnotesize{
   Illustration of the ring topology described in the main text. $\t$ 
   and  $\t_{Cu}$  
  represent  O-O and Cu-Cu links, respectively, as explained in 
  detail in Section \ref{intercell} below.}}
\label{topo}
\end{center}
\end{figure}
In the following we shall refer to the central site as Cu and to the four
external sites as O just to distinguish their position in the unit
cell. We shall see that numerical solutions of such a model clearly show
superconducting pair hopping if the total number of particles is
$2|\L|+2p$ where $|\L|$ is the number of units and $0<p<|\L|$; in
particular, 
once a magnetic field is switched on into the ring, SFQ is unambiguously
observed. 
We note that in our model no superconducting response is obtained
with less than 2 particles per 5-site unit [CuO$_{4}$]. This means that the
Zhang-Rice picture\cite{zr} for the two-dimensional $d-p$
model does not work in the present context and we wish to explore
a scenario with a larger number of particles.
The data are interpreted by implementing a
cell-perturbation theory which is  very accurate for weak
links. This low-energy approach leads to an effective
hard core boson Hamiltonian which naturally  describes itinerant
pairs and SFQ in mesoscopic rings as well. We feel that SFQ from
purely repulsive Hubbard models is interesting by itself even if it
may have no direct relevance for high-$T_{c}$ cuprates.
The tunneling of pairs has been studied in the context of the t-J model
too\cite{baskaran}, but here we wish to study symmetric Hubbard
clusters for a broad range of $U/t$ including the weak coupling limit.

The plan of the paper is the following. In Section \ref{W=0} we
recall that the $\cuoq$ cluster has two-body singlet eigenstates without
double 
occupation  called $W=0$ pairs. By exploiting a symmetry driven
configuration 
interaction mechanism they get bound once dressed by
the virtual electron-hole excitations; the binding is due
to an effective attraction among the particles of the $W=0$ pair.
In Section \ref{intercell} we introduce the model
Hamiltonian with symmetric clusters, as units of an arbitrary graph,
linked by intercell hoppings. For any given graph, we deduce a
low-energy effective Hamiltonian in Section \ref{loweneffham}. Partial
results
presented elsewhere\cite{JOPC2002} are extended to any number of
conducting pairs and we
show that the effective Hamiltonian is equivalent to the $1d$
antiferromagnetic Heisenberg-Ising model for a ring topology.
In Section \ref{sdd} we recall a recently proposed
spin-disentangled diagonalization technique that allowed us to
carry out the numerical calculations. We also give practical details on the
performance of this novel approach. Results are presented and interpreted
physically in Section \ref{2ring} for the two-unit ring and
in Section \ref{3ring} for the three-unit ring and compared
with the analytic predictions. For the larger ring, after reviewing
from Ref. \cite{JOPC2002} the flux dependence of the energy levels in
the case of O-O intercell bonds  we
analyze the current and show that it is a discontinuous function of
the flux and can be paramagnetic as well as diamagnetic.
Novel results for the case of Cu-Cu bonds are presented, and this
case is definitely more exotic:
the bound   pairs have an infinite effective mass and the flux
quantization is normal.
Our main conclusions are summarized
in Section \ref{conclusion}.

\section{$W=0$ Pairing  in Hubbard Models}
\label{W=0}

The repulsive Hubbard Hamiltonian has {\em two-body} singlet eigenstates
without double 
occupation\cite{cibal1}\cite{cibal2}\cite{cibal3}\cite{cibal4} called
$W=0$ pairs. 
Such solutions are also allowed in the fully symmetric clusters ${\cal C}$.
In the {\em many-body} ground state these pairs get dressed and
bound, and this is signaled by
$\D_{{\cal C}}(N)<0$ where $\D_{{\cal C}}(N)=E^{(0)}_{{\cal C}}(N)+
E^{(0)}_{{\cal C}}(N-2)-2E^{(0)}_{{\cal C}}(N-1)$; $E^{(0)}_{{\cal C}}(N)$
is  the interacting ground state energy of  the cluster ${\cal C}$
with $N$ particles.  By means of a non-perturbative canonical
transformation\cite{EPJB1999}\cite{SSC1999}, it can also be shown that
$\D_{{\cal C}}(N)<0$ is due to an attractive effective interaction and
at weak coupling  $|\D_{{\cal C}}(N)|$ is just the
binding energy of the pair. The extension of the theory to the full
plane was also put forth in Ref.\cite{EPJB1999}.

The $C_{4v}$ symmetric 5-site cluster is the smallest one where the
$W=0$ pairing mechanism works. We have already described
$W=0$ pairing  in great detail as a function of the one-body and interaction
parameters on all sites; the study was extended to larger clusters
too\cite{cibal3}\cite{EPJB2000}. The 5-site unit has the same topology
of the $\cu$ cluster; thus, we label the central site by the Cu
symbol and the four external ones by the O symbol. The 5-site cluster
will be also called $\cu$.
In order to 
simplify the analytical formulas,
we neglect the O-O hopping term and the only nonvanishing hopping matrix
elements are those between an O site and the central Cu site; they
are all equal to $t$. For the sake of simplicity, we parametrize
the Hubbard model in such a way that actually everything depends only on the
ratio $U/t$. Thus, we  consider the Hubbard Hamiltonian
\begin{equation}
H_{\mathrm{CuO}_{4}}=t \sum_{i\s}(
d^{\dag}_{\s}p_{i\s}+p_{i\s}^{\dag}d_{\s})+
U(\sum_{i}\hat{n}^{(p)}_{i\ua}\hat{n}^{(p)}_{i\da}+
\hat{n}^{(d)}_{\ua}\hat{n}^{(d)}_{\da})
\label{cuo4ham}
\end{equation}
where $p^{\dag}_{i\s}$ and $p_{i\s}$ are the  creation and
annihilation operators onto the O $i=1,..,4$ with spin
$\s=\ua,\da$, $d^{\dag}_{\s}$ and $d_{\s}$ are the  creation and
annihilation operators onto the Cu site,
while $\hat{n}^{(p)}_{i\s}=p^{\dag}_{i\s}p_{i\s}$ and
$\hat{n}^{(d)}_{\s}=d^{\dag}_{\s}d_{\s}$ are the
corresponding number operators. $H_{\mathrm{CuO}_{4}}$ is invariant
under the permutation group $S_{4}$. The classes of
$S_{4}$ can be labelled with the quatern $(n_{1},n_{2},n_{3},n_{4})$.
Each class contains all the permutations where
$n_{i}$ is the maximum number of sets of $i$ elements that remain
unchanged after the permutation. For example (4,0,0,0) contains all
the permutations such that 4 sets containing one element remain unchanged,
that is the identity. On the other hand (2,1,0,0) contains all the
permutations such that 2 sets containing one element and one set containing
2 
elements remain unchanged. Starting from the configuration 1234,
this class contains the permutations 1243, 1432, 1324, 4231, 3214, 2134.
$S_{4}$ has the irreducible
representations ({\em irreps}) ${\cal A}_{1}$ (total-symmetric),
${\cal B}_{2}$ (total-antisymmetric), ${\cal E}$ (self-dual),
${\cal T}_{1}$ and its dual ${\cal T}_{2}$, of dimensions
1, 1, 2, 3 and 3, respectively, see Table I. The model
admits a $W=0$ pair belonging to the irrep ${\cal E}$ and formed by
mixing degenerate one-body states.

\vspace{0.5cm}
\begin{tabular}{|c|c|c|c|c|c|}
\hline 
$S_{4}$ & $(4,0,0,0)$ & $(2,1,0,0)$ & $(0,2,0,0)$ & $(1,0,1,0)$ &
$(0,0,0,1)$  \\
\hline 
${\cal A}_{1}$ & 1 & 1 & 1 & 1 & 1 \\
\hline
${\cal B}_{2}$ & 1 & -1 & 1 & 1 & -1 \\
\hline
${\cal E}$& 2 & 0 & 2 & -1 & 0 \\
\hline
${\cal T}_{1}$ & 3 & 1 & -1 & 0 & -1 \\
\hline
${\cal T}_{2}$ & 3 & -1 & -1 & 0 & 1 \\
\hline
\end{tabular}

\vspace{0.5cm}

Table I: {\footnotesize Character Table of the $S_{4}$ group.
Each  quatern $(n_{1},n_{2},n_{3},n_{4})$
labels a class of $S_{4}$.}

\vspace{0.5cm}

For later use, we recall how the irrep $\,$
$\G \in \{ {\cal A}_{1},\;{\cal T}_{1},\;{\cal E},\;
{\cal T}_{2},\;{\cal B}_{2} \}$ of $S_{4}$ breaks in $C_{4v}$, that is the
point 
symmetry group of the square. $C_{4v}$ is a subgroup of $S_{4}$ and
its Character Table is shown in Table II. From Table I and II we have
\begin{eqnarray}
{\cal A}_{1}=A_{1},\;\;\;{\cal T}_{1}=B_{1}\oplus E,\;\;\;&&
{\cal T}_{2}=A_{2}\oplus E,\;\;\;
{\cal B}_{2}=B_{2},
\nonumber \\
{\cal E}=A_{1}&\oplus& B_{2}\;.
\label{break}
\end{eqnarray}

\vspace{0.5cm}
\begin{tabular}{|c|c|c|c|c|c|c|}
\hline 
$C_{4v}$ & $\mathbf{1}$ & $C_{2}$ & $C^{(+)}_{4},\,C^{(-)}_{4}$ &
$\s_{x},\,\s_{y}$ &
$\s_{+},\,\s_{-}$ & Symmetry \\
\hline 
$A_{1}$ & 1 & 1 & 1 & 1 & 1 & $x^{2}+y^{2}$\\
\hline 
$A_{2}$ & 1 & 1 & 1 & -1 & -1 & $(x/y)-(y/x)$\\
\hline 
$B_{1}$ & 1 & 1 & -1 & 1 & -1 & $x^{2}-y^{2}$\\
\hline 
$B_{2}$ & 1 & 1 & -1 & -1 & 1 & $xy$ \\
\hline 
$E$ & 2 & -2 & 0 & 0 & 0 & $(x,y)$ \\
\hline 
\end{tabular}

\vspace{0.5cm}

Table II: {\footnotesize Character table of the $C_{4v}$ symmetry group.
   Here  $\mathbf{1}$ denotes the identity,  $C_{2}$ the  180 degrees
   rotation,  $C_{4}^{(+)},\;C_{4}^{(-)}$ the
counterclockwise and clockwise 90 degrees rotation, $\s_{x},\;\s_{y}$
the reflection with respect to the $y=0$ and $x=0$ axis and
$\s_{+},\;\s_{-}$ 
the reflection with respect to the
$x=y$ and $x=-y$ diagonals.  In the last
column we show typical basis functions.}

\vspace{0.5cm}

>From Eq.(\ref{break}) we see that we may label the two components of
the $W=0$-pair irrep ${\cal E}$ in terms of the irreps $A_{1}$ and
$B_{2}$ of $C_{4v}$. For example, defining by
$p^{\dag}_{x,\s}=\frac{1}{\sqrt{2}}(p_{1,\s}^{\dag}-p_{3,\s}^{\dag})$ and
$p^{\dag}_{y,\s}=\frac{1}{\sqrt{2}}(p_{2,\s}^{\dag}-p_{4,\s}^{\dag})$ two
of the three degenerate eigen-operators of the kinetic term in
$H_{\cu}$, the $B_{2}$ component of
the $W=0$ pair is obtained by acting with the two-body singlet
operator 
\begin{equation}
\frac{1}{\sqrt{2}}(p^{\dag}_{x,\ua}p^{\dag}_{y,\da}+
p^{\dag}_{y,\ua}p^{\dag}_{x,\da})
\label{b2singlet}
\end{equation}
on the vacuum.

The ground state of $\cu[2]$
({\em i.e.} $\cu$ with 2 particles) belongs to $^{1}{\cal A}_{1}$ and that
of 
$\cu[4]$ is in $^{1}{\cal E}$; both are singlets, as the notation
implies.  
The ground state of $\cu[3]$ is a $^{2}{\cal T}_{1}$ doublet.
$\Delta_{\cu}(4)$ has a minimum at $U\approx 5\; t$ for this model,
as shown in Fig.(\ref{delta}), and it is negative  when
$0< U < 34.77\;t$. 
We emphasize that $\D_{\cu}(4)$ becomes positive for large values
of $U/t$ and hence pairing disappears in the strong coupling regime. In the
present problem
$U$ must exceed several tens of times $t$
before the asymptotic {\em strong coupling  regime} sets in. A
perturbation theory will strictly apply at {\em weak coupling} where
the second derivative of the curve is negative;  however,
qualitatively a weak coupling approach is rewarding in all the
physically interesting range of parameters. The above
pairing mechanism does not work in the neighborhood of
the  infinite $U$ limit.
\begin{figure}[H]
\begin{center}
   \epsfig{figure=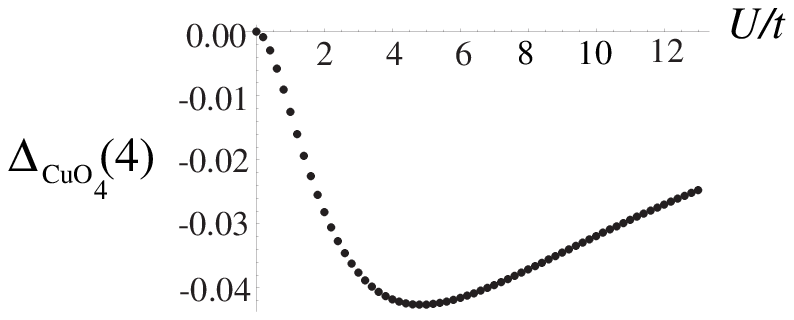,width=7cm}\caption{\footnotesize{
    $\D_{\cu}(4)$ (in $t$ units) as a function of $U/t$.  The maximum
    binding occurs at $U \sim 5 t$ where
       $\D_{\cu}(4) \approx -0.042\;t$.  For $U> 34.77\; t$ (not
       shown),
       $\D_{\cu}(4)$ becomes positive and pairing disappears.}}
\label{delta}
\end{center}
\end{figure}

We analyzed the pairing mechanism in detail in Ref.\cite{cibal4}; for the
sake of simplicity here  we report  in Fig.(\ref{diagram}) the leading,
second-order two-body amplitude for particles of opposite
spins in the degenerate $(x,y)$ orbitals of $\cuoq$.
We demonstrated\cite{cibal4}
that this produces an effective
interaction, which pushes down the singlet in Eq.(\ref{b2singlet}) and up
the triplet by $|\D_{\cu}(4)|$.  In this way, $\D_{\cu}(4)$ can be redefined
without
any reference to the ground state of clusters with a different number of
particles, and we are free from the objections based on a possible
Jahn-Teller distortion
of odd-$N$ clusters\cite{mazumdar}.

\begin{figure}[H]
\begin{center}
   \epsfig{figure=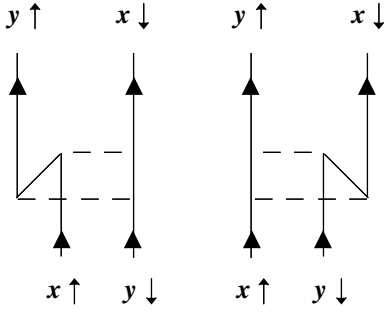,width=5cm}\caption{\footnotesize{
   The second order spin-flip diagrams for the two-body amplitude.}}
\label{diagram}
\end{center}
\end{figure}

The formation of bound pairs from purely repulsive interations was
first proposed in a pioneering paper\cite{kl} by Kohn and Luttinger.
They showed that any three-dimensional Fermi liquid undergoes a
superconducting transition by Cooper pairs of very large angular
momentum $l$. A simplified view of the Kohn-Luttinger effect is given
by considering one particle of the pair as an external charge.
Then, the screening gives rise to a long-range oscillatory potential
(Friedel oscillations) due to the singularity of the longitudinal
dielectric function at 2$k_{\rm F}$; here $k_{\rm F}$ is the Fermi
wavevector. The strict reasoning exploits the fact that the Legendre
expansion coefficients of any regular direct interaction between
particles of opposite momentum drops off exponentially in $l$. On the
other hand, the second-order contribution to the scattering amplitude
falls as $1/l^{4}$ and at least
for odd $l$ leads to an attractive interaction. In the modern
renormalization group
language\cite{shankar}, the second-order correction is obtained by
summing up the marginal scattering amplitudes of the isotropic Fermi
liquid coming from the so-called Forward channels, including, for
antiparallel 
spins, a spin-flip diagram. This scenario
does not work in the two-dimensional Fermi liquid, but going beyond
the second-order perturbation theory the Kohn-Luttinger effect is
recovered\cite{chubukov}. In the Hubbard model, there  is first-order
interaction
only for antiparallel spins, but it vanishes for $W=0$ pairs; in
second-order, in the
singlet channel, the spin-flip diagram is the only one that survives. We
found 
pairing  in  the  singlet channel in a variety of models including
carbon nanotubes\cite{psc}. Hence, the $W=0$ pairing mechanism
in the Hubbard model belongs to the broad cathegory of Kohn-Luttinger
effects with no direct interaction and a second-order correction coming from
the 
spin-flip channel only, as shown in Fig.(\ref{diagram}).

\section{Modelling Intercell Hopping }
\label{intercell}

We use  CuO$_{4}$ units as nodes of a graph $\L$.
The total Hamiltonian is
\begin{equation} 
H_{\rm tot} = H_{0}+H_{\t},
\label{hlattice}
\end{equation}
with 
\begin{eqnarray}
H_{0}=\sum_{\a\in \L} [ t \sum_{i\s}( d^{\dag}_{\a \s}p_{\a, i\s}+
p_{\a, i\s}^{\dag}d_{\a\s})+\quad\quad\quad\nonumber \\
U(\hat{n}^{(d)}_{\a \ua} \hat{n}^{(d)}_{\a \da}+\sum_{i}\hat{n}^{
(p)}_{\a, i \ua}\hat{n}^{(p)}_{\a, i\da}
)  ]     
\label{senzatau} 
\end{eqnarray}
where $p^{\dag}_{\a, i\s}$ is the  creation operator
onto the O $i=1,..,4$ of the $\a$-th cell and so on, while
$H_{\t}$ is an intercell hopping Hamiltonian.
The point symmetry group of $H_{0}$ includes $S_{4}^{|\L|}$,
with $|\L|$  the number of nodes.

There are many different ways to add an intercell hopping.
Nevertheless, to preserve the symmetry that produces
the $\D_{\mathrm{CuO}_{4}}(4)<0$ property, we take  $H_{\tau}$   invariant
under the $S_{4}$  subgroup of $S_{4}^{|\L|}$. In the following we shall
consider a 
hopping term that allows a particle in the $i$-th O site of the
$\a$-th unit to move towards the  $i$-th O site of the
$\b$-th unit with hopping integral
$\t_{\a\b}\equiv|\t_{\a\b}|e^{i\th_{\a\b}}$:
\begin{equation}
H_{\tau}=\sum_{\a,\b\in\L}\sum_{i\s}
\left[\t_{\a\b} p_{\a,i\s}^{\dag}p_{\b,i\s}+{\mathrm h.c.}\right]\; .
\label{htau}
\end{equation} 

For $N=2|\L|$ and $\tau_{\a\b}\equiv 0$, the unique  ground state consists
of 2 particles in
each CuO$_{4}$ unit. Section \ref{loweneffham} (theory) and Subsection
\ref{O-O} (numerical) are devoted to  the intercell hopping
produced by  small $|\tau_{\a\b}| \ll |\Delta_{\cu}(4)|$; to study
the propagation of  $p$    pairs
we consider a  total number
of
$N=2|\L|+2p$ particles. When $U/t$ is such that
$\D_{\mathrm{CuO}_{4}}(4)<0$, each pair prefers to lie on a single
$\cu$  and for $N=2| \L | + 2p$ the unperturbed ground state is
$2^{p}$$\times$${|\L|}\choose{p}$ times degenerate
(since $^{1}{\cal E}$ has dimension 2). On the other hand,
Subsection \ref{Cu-Cu} reports the effects of an intercell
hopping $\tau_{\rm Cu}$ between Cu sites only; this does not
break the $S_{4}^{|\Lambda |}$ symmetry and therefore its consequences
on pair propagation and SFQ are drastically different.
By this sort of models one can study the interaction of several
pairs in the same system. We are using  CuO$_{4}$ as the unit just for the
sake of 
simplicity, but  the $W=0$ mechanism produces bound pairs at
different fillings for larger clusters\cite{cibal4} too. By replacing
CuO$_{4}$ by larger units one can  model other ranges of filling
fraction.

\section{Low-energy Effective Hamiltonian for O-O Intercell Hopping}
\label{loweneffham}

In order to study the propagation of the $p$ added pairs, we obtain
an effective Hamiltonian by the cell-perturbation method
with $H_{0}$, Eq.(\ref{senzatau}), the
``cell-Hamiltonian'' and $H_{\t}$, Eq.(\ref{htau}), the
``intercell perturbation'' and by taking into account only the
low-energy singlet sector. We note that the cell-perturbation method was
already
used in Ref.\cite{fedro} and in Ref.\cite{jefferson} to support 
the original Anderson's
conjecture\cite{anderson} on the ``low-energy equivalence'' between the
$d-p$ model
(proposed by Emery\cite{emery}) and the single-band Hubbard model.
Despite the analogies with Ref.\cite{jefferson} (like the same
cell-Hamiltonian
and weak O-O links between different cells) our intercell
perturbation is different and, more important, it is the
low-energy sector which differs (one needs to consider $\cu$ units with 2, 3
and 4
bodies, in contrast with 0, 1 and 2 bodies of
Ref.\cite{jefferson}).

Let us introduce some useful notation. Let
$|\Psi^{(N)}_{0}(\a)\ket,\;\a=1\ldots |\L| $ be the ground state of
the $\a- \rm{th}$ $\cu$ unit with $2\leq N \leq 4$ particles and
$E^{(0)}(N)$ the 
corresponding energies.
Let $ {\cal S} \subset\L$, with $|{\cal S}|=p$, be the set of $p$ $\cu$
units occupied by four
particles and $\bar{{\cal S}}$ its complement in $\L$, that is
$\bar{{\cal S}} \equiv \L \setminus {\cal S}$. When $\Delta_{\cu}(4)
<0$, the unperturbed ground state
with $2| \L | + 2p$ particles can be written as
\begin{equation}
| \F_{0}^{{\cal S}} \ket = \prod_{\a \in {\cal S}}{ | \Psi_{0}^{(4)}( \a )
\ket } 
\prod_{ \b \in \bar{{\cal S}}}
 { | \Psi_{0}^{(2)}( \b ) \ket } \;
\label{pieno}
\end{equation}
and its unperturbed eigenenergy is given by
\begin{equation}
H_{0}| \F_{0}^{{\cal S}} \ket  = \left[  pE^{(0)}(4)+   ( | \L | -p
)E^{(0)}(2)  \right]
| \F_{0}^{{\cal S}} \ket \equiv
E^{(0)}_{p}| \F_{0}^{{\cal S}} \ket \; .
\end{equation}
The energies needed to excite the $\cu$ cluster from the ground state
for $U=5\div 6\;t$ [when $\D_{\cu}(4)\approx -0.04\, t$, see
Fig.(\ref{delta})] is  $\approx$ $t$  with 2 bodies and  $\approx 0.1\; t$
with 4 bodies; both are large compared with $|\D_{\cu}(4)|$ and
no level crossings take place if $|\t_{\a\b}| \ll |\D_{\cu}(4)|$.

The perturbation $H_{\t}$ in second
order will remove part of the
$2^{p}$$\times$${|\L|}\choose{p}$ degeneracy.
Let $| \F_{0} \ket $ be an exact  eigenstate with eigenenergy $E$.
Expanding $|\F_{0}\ket$   as
\begin{equation}
|\F_{0}\ket  \simeq \sum_{{\cal S}'}{a_{{\cal S}'} | \F_{0}^{{\cal S}'}
\ket} 
\end{equation}
one gets 
\begin{equation}
(E-E^{(0)}_{p})a_{{\cal S}}= \sum_{{\cal S}'}
\left[\sum_{m}\frac{  \bra \F^{{\cal S}}_{0} | H_{\t} |
\F_{m} \ket  \bra \F _{m} | H_{\t} | \F^{{\cal S}'}_{0} \ket}{E-E^{(0)}_{m}}
\right]a_{{\cal S}'} \;
\label{fssheq}
\end{equation} 
where $\left\{ |\F_{m} \ket  \right\}$ is a complete set of
excited eigenstates of $H_{0}$ and  $\{E^{(0)}_{m}\}$ their eigenenergies.

Our crucial approximation
is now to truncate the sum over the excited states
$\left\{ | \F_{m} \ket  \right\}$ considering
only the low-energy states of the form
\begin{equation}
| \F_{0}^{{\cal T},{\cal D}} \ket = \prod_{\a \in {\cal T}}
{ | \Psi^{(4)}_{0}(\a) \ket  }  \prod_{\g \in {\cal D}}
{ | \Psi^{(3)}_{0}(\g) \ket  }  \prod_{\b \in \overline
{ {\cal T} \cup {\cal D}  }  }{ | \Psi^{(2)}_{0}(\b) \ket  }   \; ,
\label{exstat}
\end{equation}
where ${\cal T}\subset \L$ is the set of $|{\cal S}|-1 \equiv |{\cal T}|$
CuO$_{4}$ 
units with 4 particles, obtained by removing one particle in one of the
previous $|{\cal S}|$ units with 4 particles; in this way we get $|{\cal
D}|=2$ 
cells with 3 particles; the remaining
$|\L|-|{\cal S}| -1 \equiv |\overline{ {\cal T} \cup {\cal D}}|$ cells have
2 particles. 
This approximation is legitimated by the fact that the first
excited state with 3 particles is $\approx$ $t$ above the ground state for
$U$ in the range 
$5\div 6$ $t$.  

The energy of the excited states in Eq.(\ref{exstat}) is
\begin{displaymath}
\tilde{E}^{(0)}_{p}= (|{\cal S}|-1)E^{(0)}(4) + 2E^{(0)}(3)+(|\L|-|{\cal
S}|-1)E^{(0)}(2) 
\end{displaymath}
and does not depend on the sets ${\cal T}$ and ${\cal D}$.
Within this approximation the Schr\"odinger equation (\ref{fssheq})
reduces to
\begin{equation}
\frac{1}{\D_{\cu}(4)}\sum_{{\cal T},{\cal D} }  \sum_{S'}
{  \bra \F^{{\cal S}}_{0} | H_{\t} |
\F^{{\cal T},{\cal D}}_{0} \ket  \bra\F^{{\cal T},{\cal D}}_{0}
| H_{\t} | \F^{{\cal S}'}_{0} \ket}\, a_{{\cal S}'} = \varepsilon  a_{{\cal
S}} \; ,
\label{Schr1}
\end{equation}
where $\ve\equiv E-E_{p}^{(0)}$ and we have disregarded contributions of
higher order in $\ve$.

The amplitude $a_{{\cal S}}\equiv a(\a_{1},\ldots,\a_{p})$ is totally
symmetric 
with respect the permutations of the distinct indices
$\a_{1},\ldots,\a_{p}$. Letting
${\cal K}(\a)=\{\b\in\L:\t_{\a\b}\neq 0\}$, after some algebra
Eq.(\ref{Schr1}) may be written in the form:
\begin{eqnarray}
\sum_{j=1}^{p}\sum_{\b\in{\cal K}(\a_{j})}
\prod_{i\neq j}(1-\d_{\b\a_{i}}){\cal J}_{\b,\a_{j}}
[a(\a_{1},\ldots,\a_{p})+
\nonumber \\ e^{2i\th_{\b\a_{j}}}
a(\a_{1},..,\a_{j-1},\b,\a_{j+1},..,\a_{p})
]=\ve a(\a_{1},\ldots,\a_{p}).
\label{schrexpl}
\end{eqnarray} 
This is a Schr\"odinger equation for $p$ hard-core bosons
with a complex effective hopping integral; below the ${\cal J}$ coefficients
will be calculated analitically and studied as a function
of the ratio $U/t$.
In Eq.(\ref{schrexpl}), the second  term in the l.h.s. describes pair
propagation, {\em e.g.} from unit $\a_{j}$ to an {\em unoccupied} unit $\b$;
in 
the first term, the system gets back to the initial state after
virtually exploring unit $\b$;  $\prod_{i\neq j}(1-\d_{\b\a_{i}})$
takes into account that if  $\b$ is  one of the
{\em occupied} units, the
particle cannot move toward it.

\subsection{Evaluation of the Effective Hopping Integral and Selection
Rules}

In Eqs.(\ref{pieno})-(\ref{exstat}) we omitted the irrep labels of
$|\Psi_{0}^{(2)}\ket$, $|\Psi_{0}^{(3)}\ket$ and $|\Psi_{0}^{(4)}\ket$  in
order to avoid a proliferation of indices. Nevertheless, to
calculate the ${\cal J}$'s we need to
reintroduce them. The ground state with
two particles is non-degenerate; it belongs to the irrep ${\cal A}_{1}$ of
$S_{4}$ that  
coincides with $A_{1}$ in $C_{4v}$, and the change in notation is:
\begin{equation}
|\Psi_{0}^{(2)}\, \ket  \longrightarrow  |\Psi_{0}^{(2),{\cal A}_{1}}\,
\ket =|\Psi_{0}^{(2),A_{1}}\,\ket   .
\end{equation}
As far as the ground state with three particles is concerned, we recall
that it is three times degenerate (apart from the trivial spin
degeneracy) and belongs to the
irrep ${\cal T}_{1}$ of $S_{4}$ that in $C_{4v}$ breaks into $B_{1}\oplus
E$:
\begin{equation}
|\Psi_{0}^{(3)}\, \ket  \longrightarrow  |\Psi_{0}^{(3),{\cal
T}_{1}^{(r)}}\, 
\ket \;\;\;\;  r =1,2,3
\end{equation}
where we may set up our basis such that ${\cal T}_{1}^{(1)}=B_{1}$ , ${\cal
T}_{1}^{(2)}=E_{x}$ and
${\cal T}_{1}^{(3)}=E_{y}$.

Finally, the ground state with four particles is two times degenerate and
belongs to the irrep ${\cal E}$ of $S_{4}$ that in $C_{4v}$ breaks
into $A_{1}\oplus B_{2}$:
\begin{equation}
|\Psi_{0}^{(4)}\, \ket  \longrightarrow  |\Psi_{0}^{(4),{\cal E}^{(s)}}\,
\ket \;\;\;\;  s =  1,2
\end{equation}  
where ${\cal E}^{(1)}= A_{1}$ and ${\cal E}^{(2)}=B_{2}$.
Useful selection rules may be obtained using group theory. By
exploiting the invariance of $H_{\t}$ under the group
$S_{4}$ and omitting, for the sake of clarity, the spin indices  in the
states 
$|\Psi_{0}^{(3)}\, \ket $, it follows that
\begin{equation}
\bra \Psi^{(4),{\cal E}^{(s)}}_{0}(\a) |
\bra \Psi^{(2),A_{1}}_{0}(\b) |   H_{\t} |
\Psi^{(3),{\cal T}_{1}^{(r_{\a})}}_{0} (\a ) \ket
|\Psi^{(3),{\cal T}_{1}^{(r_{\b})}} _{0} (\b ) \ket
\label{hopmatele} 
\end{equation}
is non-vanishing if and only if
\begin{equation}
{\cal T}_{1}^{(r_{\a})}={\cal T}_{1}^{(r_{\b})}
\label{selrule1}
\end{equation}
for ${\cal E}^{(s)}=A_{1}$; instead for ${\cal E}^{(s)}=B_{2}$,
\begin{equation}
{\cal T}_{1}^{(r_{\a})}=E_{x},\;\;
{\cal T}_{1}^{(r_{\b})}=E_{y}
\label{selrule21}
\end{equation}
or  
\begin{equation}
{\cal T}_{1}^{(r_{\a})}=E_{y},\;\;
{\cal T}_{1}^{(r_{\b})}=E_{x}.
\label{selrule22}
\end{equation}

In the case $s=1$, we have ${\cal E}^{(1)}=A_{1}$ and hence by using
the selection rules in Eq.(\ref{selrule1}),
the matrix element in Eq.(\ref{hopmatele}) consists of three
different contributions coming from the virtual 3-body states
of symmetry $B_{1}\otimes B_{1}$, $E_{x}\otimes E_{x}$ and
$E_{y}\otimes E_{y}$:
\begin{eqnarray}
\t_{\a\b}^{\rm eff}[B_{1},B_{1}]\equiv -
\t_{\a\b}\sum_{i}
\bra\Q^{(3),B_{1}}_{0,\ua}(\a)|p^{\dag}_{\a,i\ua}|
\Q^{(2),A_{1}}_{0}(\a)\ket\,\times \nonumber \\
\bra\Q^{(3),B_{1}}_{0,\da}(\b)|p_{\b,i\ua}|
\Q^{(4),A_{1}}_{0}(\b)\ket \; ;
\nonumber 
\end{eqnarray}
\begin{eqnarray}
\t_{\a\b}^{\rm eff}[E_{\m},E_{\m}]\equiv -
\t_{\a\b}\sum_{i}
\bra\Q^{(3),E_{\m}}_{0,\ua}(\a)|p^{\dag}_{\a,i\ua}|
\Q^{(2),A_{1}}_{0}(\a)\ket\,\times\nonumber \\
\bra\Q^{(3),E_{\m}}_{0,\da}(\b)|p_{\b,i\ua}|
\Q^{(4),A_{1}}_{0}(\b)\ket \; ,
\nonumber
\end{eqnarray}
with $\m=x,y$. Therefore one can solve the Schr\"odinger equation
(\ref{schrexpl}) with
\begin{equation}
{\cal J}_{\a,\b}=2\times
\frac{|\t_{\a\b}^{\rm eff}[B_{1},B_{1}]|^{2}+
|\t_{\a\b}^{\rm eff}[E_{x},E_{x}]|^{2}+
|\t_{\a\b}^{\rm eff}[E_{y},E_{y}]|^{2}}{\D_{\cu}(4)}\;,
\end{equation}
where the factor 2 comes from the spin degeneracy.
In Fig.(\ref{tauflu}) we show the
trend of ${\cal J}$ in units of $|\t|^{2}$ versus $U/t$.
The case ${\cal E}^{(2)}=B_{2}$ is
similar and may be obtained by group theory.
\begin{figure}[H]
\begin{center}
   \epsfig{figure=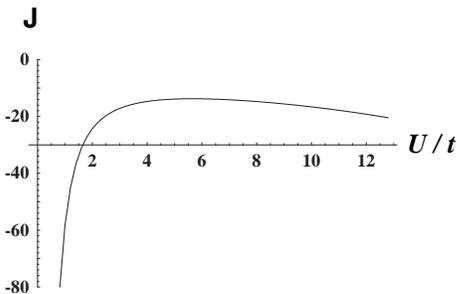,width=6cm}\caption{\footnotesize{
   ${\cal J}$ in units of $|\t|^{2}$ versus $U/t$. The formal
   divergence as $U \rightarrow 0$ is not serious since the effective
   theory holds for intermediate $U/t$, when $|\Delta_{\cu}(4)|$ is large
enough.}}
\label{tauflu}
\end{center}
\end{figure}

The above treatment holds for any $\L$; in the next Subsection we shall
specialize to the case of a one-dimensional
chain with hopping integrals $\t_{\a\b}$ only between nearest neighbors
units.

\subsection{Ring Shaped System}
\label{sup}

In order to discuss the propagation of bound pairs and the
quantization of a magnetic flux, we use a chain with periodic
boundary conditions and nearest neighbors hopping matrix elements
$\t_{\a\b}$:
\begin{equation}
\t_{\a\b}=\left\{\begin{array}{ll}
\t & {\rm if}\;\; \b=\a+1, \; \\
\t^{\ast} & {\rm if}\;\; \b=\a-1, \; \\
0 & {\rm otherwise}\;\end{array}\right.\;\quad\quad
\;\t=|\t|e^{\frac{2\pi i}{|\L|}\frac{\f}{\f_{0}}}.
\label{tau}
\end{equation}
For the sake of simplicity we first analize the case with just one added
pair ($p$=1) in detail. If the pair belongs to the
$A_{1}$ component of the ${\cal E}$ irrep  the zeroth-order
ground-state is  $|\L|$ times degenerate.
The effective  Schr\"odinger equation (\ref{schrexpl}) reads
\begin{eqnarray}
\ve a(\a)={\cal J}\left[2a(\a)+
e^{\frac{4i\p}{|\L|}\frac{\f}{\f_{0}}}
a(\a+1)+e^{-\frac{4i\p}{|\L|}\frac{\f}{\f_{0}}}a(\a-1)\right]
\nonumber 
\end{eqnarray}
which is readly solved by Fourier transforming and yields
the following eigenvalues
\begin{equation}
\varepsilon_{k}=2{\cal J}
\left[1+\cos\frac{2\p }{|\L|}\left(k+2\frac{\f}{\f_{0}}\right)\right]\;,
\;\;\;\;\;k=1,\ldots,|\L|\;.
\label{chain}
\end{equation} 
The presence of the factor 2 in front of $\f/\f_{0}$ implies
that  the model quantizes the flux in units of $\f_{0}/2$,
like superconducting pairs do. Indeed, the ground state energy
$E_{\Lambda}^{(0)}(2 |\Lambda| +2) = \min_{k}\varepsilon_{k}$ is
strictly periodic in $\phi$ with period $\phi_{0}/2$ for any
$|\Lambda|>2$. 

The case $|\Lambda| = 3$ will be used below for a numerical test of
Eq.(\ref{chain}). The lowest state energies for every quasimomentum
in the three-unit ring with 8 particles are plotted in
Fig.(\ref{fluxana}) versus $\f/\f_{0}$.
It can be shown, see below, that $|\L| = 3$ is the shortest ring
showing this effect.
\begin{figure}[H]
\begin{center}
   \epsfig{figure=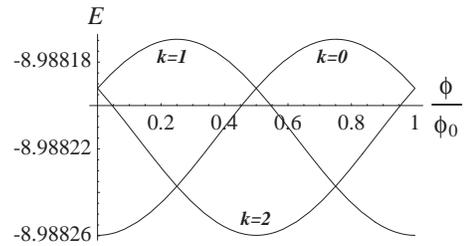,width=6cm}\caption{\footnotesize{
   Results  of Eq.(\ref{schrexpl}) for the three-unit ring  with
    $|\tau|=0.001\; t$, $U=5\; t$. Lowest-energy eigenvalues  labeled by
their 
   intercell quasi-momentum are shown versus flux $\phi$. All energies are
in $t$ 
   units.}}
\label{fluxana}
\end{center}
\end{figure}

In the case of $p$ pairs, Eq.(\ref{schrexpl}) yields
\begin{eqnarray}
\sum_{j=1}^{p}\sum_{\b=\pm 1}
\prod_{i\neq j}(1-\d_{\a_{j}+\b,\a_{i}}){\cal J}
[a(\a_{1},\ldots,\a_{p})+\quad\quad\quad
\nonumber \\  e^{\b\frac{4i\p}{|\L|}\frac{\f}{\f_{0}}}
a(\a_{1},..,\a_{j}+\b,..,\a_{p})
]=\ve a(\a_{1},\ldots,\a_{p})
\label{scring}
\end{eqnarray}  
and the model is equivalent to the Heisenberg-Ising spin chain
governed by the Hamiltonian
\begin{equation}
H_{\rm HI}=\sum_{\a=1}^{|\L|}{\cal J}
[2\eta\s^{z}_{\a}\s^{z}_{\a+1}+
e^{\frac{4i\p}{|\L|}\frac{\f}{\f_{0}}}\s^{+}_{\a+1}\s^{-}_{\a}+
e^{-\frac{4i\p}{|\L|}\frac{\f}{\f_{0}}}\s^{+}_{\a}\s^{-}_{\a+1}]
\label{heiis}
\end{equation}
where the $\s$'s are Pauli matrices, spin up represents an empty site
and spin down represents a pair. $\eta$ is the so called anisotropy
parameter and to reproduce Eq.(\ref{scring}) we must choose $\eta=-1$.
For $\eta=1$, we have the isotropic Heisenberg interaction.
By performing a Jordan-Wigner transformation, the Hamiltonian in
Eq.(\ref{heiis}) can also be mapped into a model of spinless fermions on the
ring. In the absence of a threading magnetic field ($\f=0$) the problem
was originally studied by Bloch\cite{bloch} and then exactly solved
by Hulthen\cite{hulten} [in the case $\eta=-1$] and
Orbach\cite{orbach} [in the case $\eta\leq -1$] using the Bethe's
hypothesis\cite{bethe}. A systematic analysis in the whole range of
parameters was given by Yang and Yang in a self-contained series of
papers\cite{yang}. Here we just recall that the model has a gapless
phase if $|\eta|\leq 1$, corresponding to the conducting state,
while an insulating phase sets in for $\eta<-1$. As in the 1$d$
Hubbard model, the ``magnetic perturbation'' ($\f\neq 0$) does not
spoil the integrability and the Heisenberg-Ising Hamiltonian remains
exactly solvable by the Bethe-ansatz method. Let us write an
eigenfunction of $H_{\rm HI}$ as
\begin{equation}
a(\a_{1},...,\a_{p})=\sum_{P}A_{P}e^{i\sum_{j}k_{Pj}\a_{j}}
\end{equation}
where $P$ is a permutation of the integers $1,\ldots,p$ and $A_{P}$
are $p!$ coefficients. Shastry and Sutherland\cite{shastry} have shown
that the variables $k_{j}$ are given by
\begin{equation}
|\L|k_{j}=2\p I_{j}+4\p\frac{\f}{\f_{0}}-\sum_{l\neq j}\th(k_{j},k_{l})
\label{betheans}
\end{equation}
with a phase shift
\begin{equation}
\th(k,q)=2\tan^{-1}\left[
\frac{\eta\sin[(k-q)/2]}{\cos[(k+q)/2]-\eta\cos[(k-q)/2]}\right].
\end{equation}
From Eqs.(\ref{heiis})-(\ref{betheans}) we readly see that the ground
state energy of the low-energy effective Hamiltonian $H_{\rm HI}$ is
periodic with period $\f_{0}/2$, independent of the number of added
pairs. Moreover $H_{\rm HI}$ is also the appropriate effective model 
for the strong-negative $U$ Hubbard model\cite{micnas}, for which evidence of 
superconductivity is clear\cite{aligia}. 
Thus we conclude that the purely repulsive $\cu$-Hubbard ring
threaded by a magnetic field quantizes the flux in a superconducting
fashion if the number of particles is $2|\L|+2p$ with $0\leq p\leq
|\L|$.
   
\section{Spin-Disentangled Diagonalization}
\label{sdd}

In order to find the lowest energy eigenvalues and eigenfunctions  when the 
dimension ${\cal N}$ of the Hilbert space is very large  one must  
avoid storing the ${\cal N} \times {\cal N} $ Hamiltonian matrix 
$ {\cal H}_{{\cal N} \times {\cal N}} $.  The obvious  recipe for sparse 
matrices, which are best handled by the Lanczos method,
prescribes storing the nonzero elements and their row and column 
addresses in  arrays. However for big problems this process is slow, 
since the matrix elements must be referenced to their position in ${\cal 
H}_{{\cal N} \times {\cal N}}$ and
must be retrieved each time for later use. One can try to improve the 
situation by projecting on the irreps of the symmetry group in order 
to reduce ${\cal N}$; however this involves building the projected 
basis and then  new  Hamiltonian matrix elements, and the new 
Hamiltonian is much less sparse than before, which destroys much of 
the benefit gained by the symmetry.

Here we   exactly diagonalize the $|\L|=2$ and $|\L|=3$ ring Hamiltonian 
by  the {\em Spin-Disentangled}  technique,
which we briefly introduced recently\cite{JOPC2002}, but deserves a
fuller illustration.  This  allows to solve the  ${\cal N} \times {\cal 
N} $ many-electron problem
by storing  and handling  $\sqrt{{\cal N}} \times \sqrt{{\cal N}} $  
matrices. One of its  clear advantages is that it  {\em does not require the Hamiltonian to be sparse}.
The method is also much 
more convenient than the usual one, since the $\sqrt{{\cal N}} \times \sqrt{{\cal N}} $ matrices are directly used 
by matrix multiplication, which is a fast process, and no mapping back 
onto ${\cal H}_{{\cal N} \times {\cal N}}$ is involved. All the 
advantages of symmetry are gained simply by using a projected 
starting state of the Lanczos chain.
To the best of our
knowledge, the {\em Spin-Disentangled}  technique was not invented earlier, which is somewhat
surprising, being a very general method.

We let  $M_{\ua}+M_{\da}=N$ where $M_{\s}$ is the number of  particles
of spin $\s$; $\{|\f_{\a\s}\ket\}$ is a real orthonormal basis, that
is, each vector is a homogeneous polynomial in the $p^{\dag}$ and $d^{\dag}$
of degree $M_{\s}$ acting on the vacuum. We write the
ground state wave function in the form
\begin{equation}
|\Psi\rangle=\sum_{\alpha \beta}
L_{\alpha \beta}|\phi_{\alpha \uparrow}\rangle
\otimes |\phi_{\beta \downarrow} \rangle
\label{lali}
\end{equation}
which shows how the $\ua$ and $\da$ configurations are entangled. The
particles of one spin are treated as the ``bath'' for those of the
opposite spin: this form also enters the proof of a famous theorem by
Lieb\cite{lieb}.
In Eq.(\ref{lali})   $L_{\a\b}$ is a
$m_{\ua}\times m_{\da}$ rectangular matrix with
$m_{\s}$=${5|\L|}\choose{M_{\s}}$.
We let  $K_{\s}$ denote the kinetic energy $m_{\s}\times m_{\s}$
square matrix of $H_{\rm tot}$ in the basis $\{|\f_{\a\s}\ket\}$,
and $N^{(\s)}_s$ the spin-$\s$ occupation number matrix at site
$s$ in the same basis ($N^{(\s)}_s$ is a  symmetric matrix since the
$|\f_{\a\s}\ket$'s are real). Then, $L$ is acted upon by the Hamiltonian
$H_{\rm tot}$ according to the rule
\begin{equation}
H_{\rm tot}[L]=[K_{\ua}L + L K_{\da}]+
U \sum_{s}  N^{(\ua)}_s L N^{(\da)}_s\;.
\label{lieb3}
\end{equation}
In particular for $M_{\ua}=M_{\da}$ ($S_{z}=0$ sector)
it holds $K_{\ua}=K_{\da}$ and $N^{(\ua)}_s=N^{(\da)}_s$. Thus, the action
of $H$ is obtained in a spin-disentangled way.
The generality of the method
is not spoiled by the fact that it is fastest in the $S_{z}=0$
sector, because it is useful provided that the spins are not totally
lined up; on the other hand, $S_{z}=0$ can always be assumed,
as long as the Hamiltonian is $SU(2)$ invariant.

For illustration, consider the Hubbard model with two sites $a$ and
$b$ and two electrons (H$_{2}$ molecule) each in the $\f_{a}$ or
$\f_{b}$ orbital. The intersite hopping is $t$
and the on-site repulsion $U$. In the standard method, one
sets up basis vectors for the $S_{z}=0$ sector
\begin{eqnarray}
|\q_{1}\ket=|\f_{a\ua}\ket\otimes |\f_{a\da}\ket,
\quad\quad
|\q_{2}\ket=|\f_{a\ua}\ket\otimes |\f_{b\da}\ket,
\nonumber\\ 
|\q_{3}\ket=|\f_{b\ua}\ket\otimes |\f_{a\da}\ket,
\quad\quad
|\q_{4}\ket=|\f_{b\ua}\ket\otimes |\f_{b\da}\ket,
\nonumber
\end{eqnarray}
One then looks for eigenstates (three singlets and one triplet)
\begin{equation}
|\Psi\ket=\sum_{i=1}^{4} \psi_{i}|\q_{i}\ket
\label{dard}
\end{equation}
of the Hamiltonian 
\begin{equation}
H_{{\rm H}_{2}}=\left(\begin{array}{rrrr}
U & t & t & 0 \\
t & 0 & 0 & t \\
t & 0 & 0 & t \\
0 & t & t & U \end{array}\right).\label{stand}
\end{equation}
Insted of  working with $4\times 4$ matrices, we can cope with $2\times
2$ by the spin-disentangled method using the form in Eq.(\ref{lali}) with
\begin{eqnarray*}
L=\left(\begin{array}{rr}
\psi_{1} & \psi_{2}\\
\psi_{3} & \psi_{4}  \end{array}\right),
\quad\quad K_{\s}=\left(\begin{array}{rr}
0 & t\\
t & 0 \end{array}\right), \nonumber \\
N_{a}^{(\s)}=\left(\begin{array}{rr}
1 & 0\\
0 & 0  \end{array}\right),
\quad\quad 
N_{b}^{(\s)}=\left(\begin{array}{rr}
0 & 0\\
0 & 1  \end{array}\right).\\
\end{eqnarray*}
Using Eq.(\ref{lieb3}), one finds
\begin{equation}
H_{{\rm H}_{2}}|\Psi\rangle=\sum_{\alpha=a,b}\sum_{\beta=a,b}
(H_{{\rm H}_{2}}[L])_{\alpha \beta}|\phi_{\alpha \uparrow}\rangle
\otimes |\phi_{\beta \downarrow} \rangle
\label{hlali}
\end{equation}
with
\begin{equation}
H_{{\rm H}_{2}}[L]=\left(\begin{array}{cc}
U\psi_{1}+t(\psi_{2}+\psi_{3}) & t(\psi_{1}+\psi_{4})\\
t(\psi_{1}+\psi_{4}) & U\psi_{4}+t(\psi_{2}+\psi_{3})
\end{array}\right).\\
\end{equation}
The reader can readily verify that this is the same as
applying $H_{{\rm H}_{2}}$ in the form of Eq.(\ref{stand}) to the standard
wave 
function in Eq.(\ref{dard})
and then casting the result in the form of Eq.(\ref{lali}).
Since we can apply $H_{{\rm H}_{2}}$ we can also diagonalize it.

 ${\cal N} = 4 $  for the H$_{2}$ toy model, but  in the   $S_{z}=0$ sector 
 for the $|\L|$=3 ring, ${\cal N} = 1863225$ and $\sqrt{{\cal N}} = 1365$, 
 which is a clear advantage. Here we have implemented this method for the Hubbard Hamiltonian.
We emphasize, however, that this approach will be generally useful for
the many-fermion problem, even with a realistic Coulomb interaction,
which can be suitably discretized.

\subsection{The practical numerical recipe}

We put   a symmetry-adapted trial wave function in the form in 
Eq.(\ref{lali}) and operate  the Hamiltonian matrix 
by Eq.(\ref{lieb3}); each new application introduces  a new Lanczos {\em site} 
and we can proceed by generating a Lanczos
{\em chain}. 
To this end we need  to orthogonalize to the previous {\em sites} by the
scalar product  given by $\bra \Q_{1}|\Q_{2} \ket =
{\mathrm Tr}(L_{1}^{\dag}L_{2})$.
In this way  we put  the Hamiltonian matrix in a tri-diagonal form. This
method 
is well suited since  we are mainly interested in  the
low-lying part of the spectrum.
A severe numerical instability sets in when the chain exceedes a few
tens of {\em sites}, {\em i.e.}  well before the Lanczos method
converges. Therefore we use
repeated two-site chains alternated with moderate-size ones.

In the basis of the sites (of the original cluster) the
occupation matrices $N^{(\s)}_{s}$ are diagonal with elements equal to
0 or 1, simplifying the calculation of the interaction term.
Moreover, in choosing the  trial wave function for small $\tau$
we  take full advantage from
our knowledge of the $S_{4}$ irrep of the $\tau=0$ ground state. This
speeds the calculation  by a factor of the order of 2 or 3 compared to a
random 
starting state (or even more, if $U$ is large).
Typically, starting from a  $\tau=0$ ground state for the three-unit
ring, 24 short  Lanczos {\em chains}
were enough to obtain a roughly correct energy and a 20-{\em
site chain}  achieved an accurate eigenvalue and an already stabilized
eigenvector. In limiting cases when the results could
be checked against analytic ones, using double precision routines an
accuracy better than 12 significant digits was readily obtained.
On a personal computer with a Celeron CPU the three-unit ring with
8 particles required 40 minutes for the eigenvalue, and an accurate
wave function required less than a hour.

\section{Numerical Results and Discussion: the Two-Unit ``Ring''}
\label{2ring}

In this section we analyze the two-unit ring with $\D_{\cu}(4)<0$.
Here a complex $\tau$ is  equivalent to a real one, {\em i.e.}
no magnetic field can thread the system. This is peculiar of
the ring with two clusters, since each $\cu$ is on the right as well
as on the left of the other. Indeed, we have
\begin{eqnarray}
H_{\tau}&=&(\t+\t^{\ast})\sum_{j \s}
\left(p_{1,j\s}^{\dag} p_{2,j\s} +
p_{2,j\s}^{\dag} p_{1,j\s} \right)=
\nonumber \\ &=&
2|\tau|\cos\left({\frac{\pi \phi}{\phi_0}}\right)\sum_{j \s}
\left(p_{1,j\s}^{\dag} p_{2,j\s} +
p_{2,j\s}^{\dag} p_{1,j\s} \right);
\label{2ringht}
\end{eqnarray}
hence, the ground state energy $E^{(0)}(\f,|\t|)$ as a function
of the flux $\f$ and of the modulus of $\t$ satisfies
\begin{equation}
E^{(0)}(\f,|\t|)=E^{(0)}(0,|\tau|\cos{\frac{\pi \phi}{\phi_0}}).
\end{equation}
$E^{(0)}(\f,|\t|)$ has a local maximum at $\f=\f_{0}/2$ (a property
which is independent of the number of particles in the two-unit
ring) since $H_{\tau}$ in Eq.(\ref{2ringht}) vanishes; see
Fig.(\ref{2ringene})
for the case of $2|\L|+2=6$ particles.

The translational invariance allows to label any state
by the {\em crystal momentum} $p\equiv \p\hbar k$, $k=0,1$. We introduce
the short-hand notation $\KET{+}$ and $\KET{-}$ for the
components of the 6-body non-interacting ground-state
multiplet with $k=0$ and $k=1$ respectively
\begin{equation}
\KET{\pm}=\frac{1}{\sqrt{2}}\left[\kst{4}{1}{0}\kst{2}{2}{0} \pm
\kst{2}{1}{0}\kst{4}{2}{0}\right] .
\end{equation} 
We have 
\begin{equation} 
{\bgst{3}{1}\bgst{3}{2}}{H_{\tau}}\KET{-} = 0
\end{equation}
(the singlet projection of $\kst{3}{1}{0}\kst{3}{2}{0}$ has
$k=0$ quantum number) and there is no second order correction
in the $k=1$ subspace as shown in Fig.(\ref{2ringene}).
In the $k=0$ subspace the correction is proportional to
$|\t|^{2}/\D_{\cuoq}$ for small $|\t|$,
in agreement with the analytical predictions.
As shown in Fig.(\ref{2ringene}), the maximum at
$\phi=\phi_{0}/2$ is not a cusp, as there no
level crossing is found.
\begin{figure}[H]
\begin{center}
\epsfig{figure=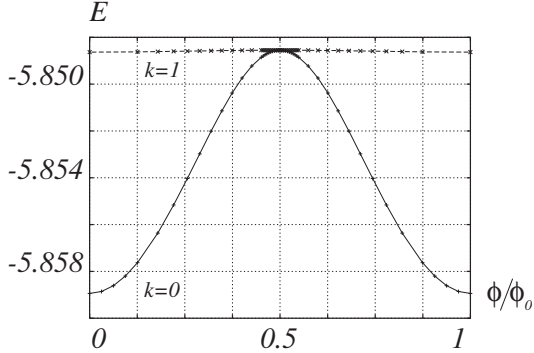,width=7cm}\caption{\footnotesize{
Energy of the ground state ($k=0$) and of the first excited state
($k=1$) with $2|\L|+2=6$ particles as a function of $\phi/\f_{0}$. Here
$U=5 t$, ($\D_{\cu}(4) \approx -0.04258\, t$) and
$|\tau|=0.001 \,t$. The energies are in units of $t$.}}
\label{2ringene}
\end{center}
\end{figure}

\section{Numerical Results and discussion: the Three-Unit Ring}
\label{3ring}

\subsection{O-O intercell hopping}
\label{O-O}

In this section we consider the three-unit ring focusing the
attention on the case $\Delta_{\cuoq}(4)<0$ and total number of
particles $2|\L|+2=8$.
The switching on of the hopping
$\tau$ between the O sites breaks the symmetry  group
$C_{3v}\otimes S_{4}^{3}$ into $C_{3v} \otimes
S_{4}$ for real $\tau$; in a magnetic field (complex $\tau$), this
further breaks into  $C_3 \otimes S_{4}$.
Real $\tau$  lifts the degeneracy between the $k=0$ subspace and the
subspaces  
$k=1$ and $k=2$ of $C_{3}$ (as usual $k$ is related to the crystal
momentum $p\equiv 2\p\hbar k/3$ in this case), but cannot split
$k=1$ and 2 because they belong to the degenerate irrep of $C_{3v}$; complex
$\tau$  resolves this degeneracy.
\begin{figure}[H]
\begin{center}
\epsfig{figure=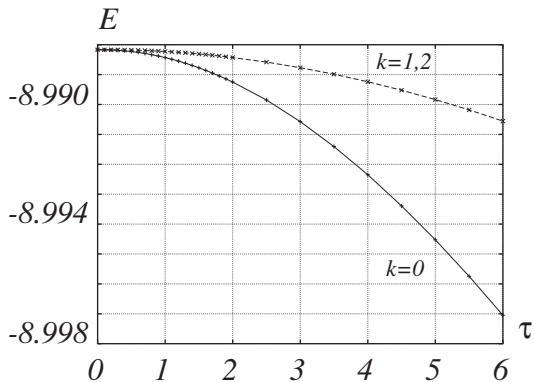,width=7cm}\caption{\footnotesize{
Energies of the ground state ($k=0$) and of the first
excited state ($k=1$ and 2) of the three-unit ring with $8$ particles
versus $|\t|$ for $U=5 t$. The energies
are in units of $t$, the parameter $|\tau|$ is in units of $10^{-3} t$.}}
\label{plot2}
\end{center}
\end{figure}

In Fig.(\ref{plot2}) we report  the numerically exact results for a ring
with 
three clusters, with  $U=5 t$  and $|\tau|$ in the
range from $0$ to $0.006 t$. The ground
state energy and the first excited level depend quadratically on $|\t|$
for small $|\t|$. As expected from Eq.(\ref{chain}), the lower eigenspace
has $k=0$, while the
first excited one contains the states with $k=1$ and $2$. Differently
from the two-unit ring, the first excited level receives
a second-order correction.

\begin{figure}[H]
\begin{center}
       \epsfig{figure=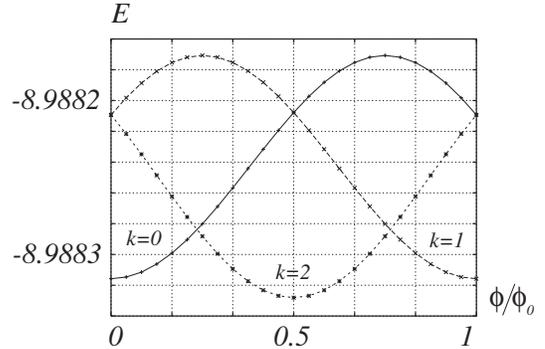,width=7cm}
       \caption{\footnotesize{Numerical results for the low energy states of
   the three-unit ring , as
a function of the concatenated magnetic flux. Here $U=5\
t$, ($\D_{\cu}(4) \approx -0.04258 \,t$) $|\tau|=0.001 \,t$.
The energy is in units of $t$.}}
\label{plot3}
\end{center}
\end{figure}

The three-unit ring is the smallest ring where we can insert a
magnetic flux $\phi$ by  $\tau=|\tau| e^{i\theta}$,
$\theta=\frac{2 \pi}{3}(\phi/\phi_0)$. The energies of the three
ground-state multiplet components are reported in Fig.(\ref{plot3})
for $|\tau|\ll |\D_{\cu}(4)|$ and $U=5t$.
At $\f=0$ the ground state belongs to the $k=0$ subspace, while
the first excited levels have $k=1$ and $2$. Their spatial degeneration
is fully lifted: the $k=1$ level increases while the $k=2$ level
decreases up to  $\phi=\phi_{0}/2$.
As $\phi$ increases, the ground state energy grows quadratically in
$\phi$ (diamagnetic behaviour). Near $\phi=\phi_{0}/4$ we
find a level crossing between  $k=0$ and $k=2$, while at
$\phi=\phi_{0}/2$, $k=0$ becomes degenerate with $k=1$ and the ground
state energy is in a new minimum belonging to the $k=2$ subspace:
a sort of ``restoring'' of the $\phi=0$ situation is taking place as
in the BCS theory\cite{lipa}. Indeed, at
$\phi=\phi_{0}/2$ the symmetry group is $\tilde{C}_{3v}\otimes
S_{4}$  where  $\tilde{C}_{3v}$ is isomorphous to
$C_{3v}$ (reflections $\s$ are replaced by $\s g$, where $g$ is a suitable
gauge transformation). This feature was also  found in other
geometries\cite{EPJB1999}\cite{EPJB2000}.
In the region from $\phi=\phi_{0}/2$ and $\phi=\phi_{0}$
we numerically verified that $E_{k=2}(\phi)=E_{k=2}(\phi_{0}-\phi)$,
$E_{k=0}(\phi)=E_{k=1}(\phi_{0}-\phi)$ and
$E_{k=1}(\phi)=E_{k=0}(\phi_{0}-\phi)$, where $E_{k}(\f)$ is the
ground state energy in the $k$ sector. The comparison of the
numerical results shown in Fig.(\ref{plot3}) with the analytic ones in
Fig.(\ref{fluxana}) supports the accuracy of the cell-perturbation scheme
proposed in Section \ref{loweneffham}. Thus, the dressed $W=0$ pair
screens the vector potential as a particle with an effective charge
$e^{\ast}=2e$ does. At both minima of $E^{(0)}(\f)$ we have
computed $\D_{3-{\rm unit}}(8)\approx -10^{-2} t$.
Here, the half-integer AB effect is actually SFQ.

Fulfilling the conditions $\Delta_{\cuoq}(4)<0$ and $|\tau| \ll
|\Delta_{\cuoq}(4)|$, we varied $U$ and $|\tau|$ and found
analogous trends for the ground state energy.
Increasing $|\tau|$ with fixed $\Delta_{\cuoq}(4)$
lowers the central minumum and depresses  the two maxima. On the
other hand, if $|\Delta_{\cuoq}(4)|$ decreases at fixed $|\tau|$ the central
minimum and the side peaks are affected in a similar way.
This is reasonable since the perturbative
parameter is $|\tau|/|\Delta_{\cuoq}(4)|$.

The three-unit ring also enables us to study  persistent diamagnetic
currents carried by bound pairs screening the magnetic flux.
We calculated  the expectation value for each $k$ of the total current
operator as a function of the flux. The current operator\cite{kohn}
\begin{equation}
\hat{I} = c \frac{\de H_{\rm tot}}{\de \phi}= \frac{e}{\hbar |\Lambda|}
\sum_{i,\a,\s}i (\tau\, p_{\a+1,i\sigma}^{\dagger} p_{\a,i\sigma} -
\tau^{*} p_{\a,i\sigma}^{\dagger} p_{\a+1,i\sigma} )
\label{current}
\end{equation}
yields a gauge invariant average $I$. By expanding $\hat{I}$ in
Eq.(\ref{current}) in
powers of $\f$ near $\f=0$ one may identify the paramagnetic and the
diamagnetic contributions with the zeroth and the first order terms
respectively\cite{dagotto}. The results are reported in
Fig.(\ref{plot4}); the current is proportional to the flux derivative
of the ground-state energy [see Fig.(\ref{plot3})] according to
the Hellmann-Feynman theorem. Near $\f=0$ the system generates a
diamagnetic current which screens the threaded magnetic field.
When $\phi$ exceedes a critical value $\sim \phi_{0}/4$, a
breakdown of the ground state occurs. This corresponds to a
discontinuity of the current which changes sign; then the current
enhances the external field.  At
$\f=\f_{0}/2$ the current vanishes again. Indeed, like at $\f=0$, the
eigenfuctions 
may be choosen real [$H_{\t}$ at $\f=\f_{0}/2$ is obtained from
$H_{\t}$ at $\f=0$ by reversing the sign of four O-O bonds connecting
two nearest neighbours units]. Thus, near $\f_{0}/2$ the magnetic flux
is still a small perturbation with respect to a new real intercell
hopping Hamiltonian and the current correctly screens the new magnetic
field.  From Fig.(\ref{plot4}) we see that the maximum value of the
diamagnetic current is of the order of $1\div 10$ nano Ampere if $t=1$
eV and the ratio $I/(\f/\f_{0})\approx e|\t|/h$ near $\f=0$.

\begin{figure}[H]
\begin{center}
       \epsfig{figure=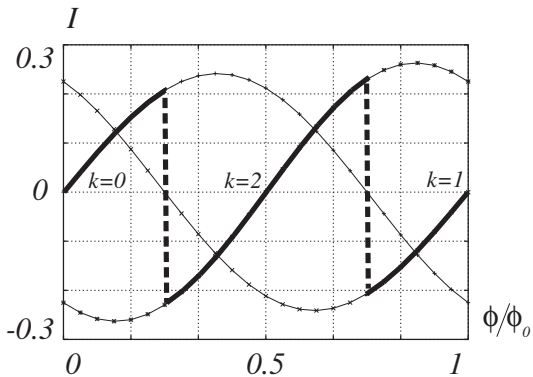,width=7cm}
       \caption{\footnotesize{Total current for the three-$\cuoq$ ring, as
a function of the magnetic flux. Here $U=5\
t$, $|\tau|=0.001 \, t$. The current is in units of $e |\tau|/h$.
The thick line marks the ground state current.}}
\label{plot4}
\end{center}
\end{figure}

In Fig.(\ref{plot5}) we show the trend of the ground state energy in
each $k$ sector for the non-interacting ($U=0$) three-unit
ring. In this case there is no pairing in $\cu$ and
indeed the ground state energy is linear in the field at small fields
(normal Zeeman effect). The lowest state is $k=2$ throughout. Interestingly,
the three-unit ring {\em concatenated with half a flux quantum} would be
diamagnetic, 
but the absence of a second minimum shows that it would be Larmor
diamagnetism. The absence of SFQ in Fig.(\ref{plot5})
is a further evidence of the repulsion-driven pairing mechanism
discussed in Section \ref{W=0}.
\begin{figure}[H]
\begin{center}
       \epsfig{figure=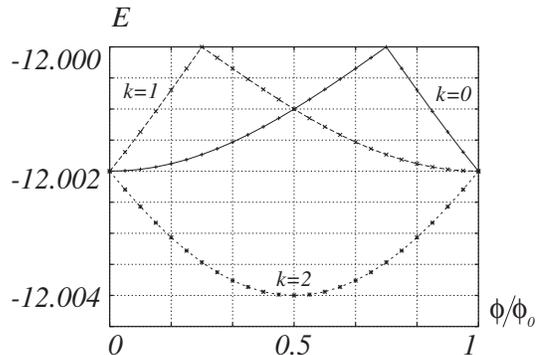,width=7cm}
       \caption{\footnotesize{Low energy states of the three-unit
       ring versus $\f/\f_{0}$. Here $U=0$ and $|\tau|=0.001\, t$.
   The energy is in units of $t$.}}
\label{plot5}
\end{center}
\end{figure}

\subsection{Cu-Cu intercell hopping}
\label{Cu-Cu}

We can alternatively model the three-unit ring
by connecting only the central Cu sites of the constituent $\cu$
units with a hopping term $\t_{\rm Cu}$; in order to
study the propagation of a bound pair we again assume the total number of
particles $2|\L|+2=8$. The full system threaded by the flux has a
$C_{3}\otimes S_4^3$ symmetry because the O sites are
not involved in the intercell Hamiltonian.
Again, we consider the case $\Delta_{\cuoq}(4)<0$ and ask  if
now the $W=0$ bound pair can screen out the magnetic flux.
\begin{figure}[H]
\begin{center}
       \epsfig{figure=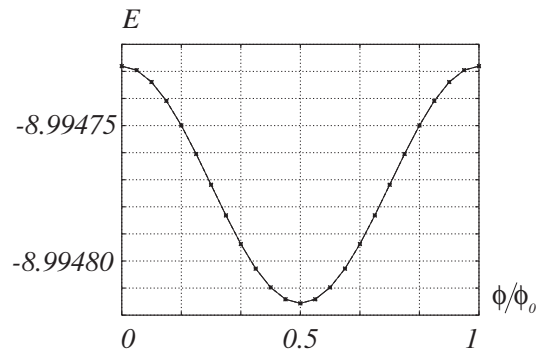,width=7cm}
       \caption{\footnotesize{ Ground state energy $E$ of
   the three-unit ring in units of $t$, as
a function of the concatenated magnetic flux. $E$ is $k$-independent
(see text). Here $U=5\
t$, $|\tau_{\rm Cu}|=0.1 t$.}}
\label{plot7}
\end{center}
\end{figure}

We find that $\tau_{\rm Cu}$ produces much smaller effects than $\t$; for
$|\tau_{\rm Cu}|\ll |\Delta_{\cuoq}(4)|$ the energy
eigenvalues are  outright  $\phi$ independent  to great accuracy.
Therefore we considered  $|\tau_{\rm Cu}|=0.1 t$; still
the dependence of the ground state energy is weak, see Fig.(\ref{plot7}).
At $\phi=0$, the correction\cite{numeri} due to $|\tau_{\rm Cu}|$ to ground
state energy is $\sim 10^{-3} t$. Moreover, remarkably,
the system behaves as a paramagnet.

The reason of this unusual behavior is the following.
There is  no flux-induced
splitting of the three $k$ levels because the $W=0$ pair is strictly
localized by the {\em local} symmetry.  Indeed the $S_{4}$ label of each
$\cu$ unit is a good quantum number. 
No SFQ is observed because the screening
of the magnetic field by the bound pair
is forbidden. The small correction to the ground state energy
comes from a second-order process. Starting {\em e.g.} with an unperturbed
state
$\KET{4,2,2}$, in which the bound pair is localized on the first
cluster, the correction involves virtual
states $\KET{3,3,2}$ and $\KET{3,2,3}$ in which a totalsymmetric
particle  
jumps forth and back on the nearby clusters.
This process occurs with a small amplitude because of a severe
energy misfit. This is particularly clear at weak coupling, when
the lowest-energy $A_{1}$ particle of the first cluster must hop to
antibonding  $A_{1}$ orbitals of the nearby clusters; the amplitude of
this process is further reduced by the overlap of these orbitals
with the localized Cu one.
However, such  virtual processes are insensitive to the flux.
Any $\phi$ dependence arises from third-order corrections (order 
$|\Lambda|$ in general). Indeed, 
the $A_{1}$ particle must go virtually  around the trip clockwise or
anticlockwise. 
In the ground state, of course, it chooses the wise in such a way to
gain energy from the magnetic field. This is why a paramagnetic
dependence on the flux is seen in Fig.(\ref{plot7}) and the correction goes
like $-\phi^{2}$ at small $\phi$.
This is interesting because it shows
how the local symmetry can hinder the tunneling of bound pairs carrying
conserved  quantum numbers; SFQ is not a necessary consequence
of superconductivity if the pairs are not totalsymmetric.

\section{Conclusions}
\label{conclusion}

We propose a  Hubbard model with on-site repulsion defined on a graph $\L$
with  
5-site $C_{4v}$-symmetric clusters as nodes in order to study the
response of the system to a threading magnetic flux. For ring-shaped
systems and weak O-O links we find a half-integer AB effect which is
unambiguosly 
interpreted as SFQ. The key ingredient is the $W=0$ pair which is a
two-body singlet eigenstate of $H_{\cu}$ without double occupation
and is formed by mixing degenerate one-body states. In the interacting
problem the $W=0$ pair becomes a bound pair when four particles lie in the
$\cu$ cluster. The pairing mechanism is due to an effective
attractive interaction mediated by repeated electron-hole exchanges with
the {\em Fermi sea}. Thus, SFQ may be found in purely repulsive 1$d$ Hubbard
models if the nodes are represented by a non-trivial basis.
Focussing on the low-energy sector, we find a simplified description of the
model 
in terms of an effective hard-core boson Hamiltonian that can be
solved exactly for  ring-shaped systems and arbitrary filling.
Further, we show that the  boson Hamiltonian is equivalent to the
well-known\cite{hulten} Heisenberg-Ising spin chain with an
antiferromagnetic anisotropy
parameter $\eta=-1$.

The analytic results are well confirmed by the
numerical findings for the two- and three-unit ring (14,400 and
1,863,225 configurations). To this end, we have recently introduced a
new exact-diagonalization technique. By disentangling  spin-up
and spin-down  we reduce the size of the matrices that must be
handled considerably; for spin-unpolarized systems the matrix
dimension is the square root of the overall size of the Hilbert space.
The method allows large further reductions by exploiting the
symmetry; this can be done in several ways. One can first diagonalize
Dirac's characters and then apply the spin-disentangled technique with
a smaller function space; alternatively, one can  set up the
spin-disentangled technique by a variational approach, a projection
operator ($H^{n}$)  or  Lanczos method, starting with
a trial state belonging to a well defined symmetry.

We have computed the ground state energy and the induced supercurrent
as a function of the trapped flux in the case of weak O-O links. We
have also studied the effect of direct intercell Cu-Cu links; in this
case, bound pair propagation is hindered by symmetry, because each unit
must keep its own $S_{4}$ irrep. Hence, the unusual situation arises
when the threading flux is not screened by the superconducting pairs
and a paramagnetic response  prevails.

}

\end{document}